\PassOptionsToPackage{hyphens}{url}
\documentclass[runningheads]{llncs}

\usepackage[letterspace=-40]{microtype} \usepackage{amssymb}
\usepackage{graphicx}
\usepackage{booktabs}
\usepackage{multirow}
\usepackage[abbreviations]{foreign}
\usepackage[inline]{enumitem}
\usepackage{hyperref}
\usepackage{amsmath}
\usepackage{cleveref}\usepackage{pifont}
\usepackage{subcaption}
\usepackage[subtle]{savetrees}
\usepackage[margin=0pt,skip=6pt,belowskip=0pt,font={small,stretch=1.1},labelfont=bf,justification=centering]{caption}
\usepackage{tabularx, makecell, booktabs, array}
\usepackage[prologue,usenames,table,dvipsnames,x11names]{xcolor}
\usepackage{tikz}
\usepackage[
  backend=biber,
  doi=false,
  style=lncs,
  minnames=3,
  maxnames=3
]{biblatex}
\usepackage[export]{adjustbox}
\usepackage{eso-pic}
\usepackage[most]{tcolorbox}
\usepackage{orcidlink}

\usepackage{hyperxmp}
\hypersetup{
    pdfauthor={Jämes Ménétrey, Christian Göttel, Anum Khurshid, Marcelo Pasin, Pascal Felber, Valerio Schiavoni, Shahid Raza},
    pdftitle={Attestation Mechanisms for Trusted Execution Environments Demystified},
}

\let\llncssubparagraph\subparagraph
\let\subparagraph\paragraph
\usepackage[compact]{titlesec}
\let\subparagraph\llncssubparagraph

\hypersetup{
  colorlinks = true, urlcolor   = blue, linkcolor  = blue, citecolor  = red   }

\DeclareFieldFormat{titlecase}{\MakeSentenceCase*{#1}}
\AtEveryBibitem{\clearfield{pages}\clearlist{location}\clearfield{pagetotal}\ifentrytype{online}{}{\clearfield{url}}\ifentrytype{inproceedings}{\clearfield{booktitle}\clearfield{year}}{}}

\usepackage{xpatch}

\makeatletter
\patchcmd\blx@bblinput{\blx@blxinit}
                      {\blx@blxinit
                      }{}{\fail}
\makeatother

\def\isExtendedCiteEnabled{0}

\ifnum\isExtendedCiteEnabled=1
  \newcommand{\extcite}[2][null]{~\cite{#2}}\else
  \newcommand{\extcite}[2][null]{\def\paramOne{#1}\def\paramNull{null}\ifx\paramOne\paramNull \else ~\cite{#1}\fi }
\fi

\newcommand{\sotaAdj}{state-of-the-art\xspace}
\newcommand{\sotaNoun}{state of the art\xspace}

\newcolumntype{R}{>{\raggedleft\arraybackslash}X}
\newcolumntype{Y}{>{\centering\arraybackslash}X}
\newcolumntype{P}[1]{>{\raggedleft\arraybackslash}p{#1}}
\newcolumntype{q}[1]{>{\centering\arraybackslash\hspace{0pt}}p{#1}}

\definecolor{prioritycolor}{HTML}{969bce}
\newcommand*{\priority}[1]{\raisebox{-1pt}{\begin{tikzpicture}[scale=0.15]\draw (0,0) circle (1);
  \fill[fill opacity=1,fill=prioritycolor] (0,0) -- (90:1) arc (90:90-#1*3.6:1) -- cycle;
  \end{tikzpicture}}}

\newcommand{\compfull}{\priority{100}}
\newcommand{\comppart}{\priority{50}}
\newcommand{\compnone}{\priority{0}}

\AtBeginDocument{\providecommand\BibTeX{{\normalfont B\kern-0.5em{\scshape i\kern-0.25em b}\kern-0.8em\TeX}}}

\makeatletter
\begin{document}

\title{Attestation Mechanisms for Trusted Execution Environments Demystified}

\author{Jämes Ménétrey\inst{1}\orcidlink{0000-0003-2470-2827} \and
Christian Göttel\inst{1}\orcidlink{0000-0002-4465-6197} \and
Anum Khurshid\inst{2}\orcidlink{0000-0002-5788-069X} \and
Marcelo Pasin\inst{1}\orcidlink{0000-0002-3064-5315} \and
Pascal Felber\inst{1}\orcidlink{0000-0003-1574-6721} \and
Valerio Schiavoni\inst{1}\orcidlink{0000-0003-1493-6603} \and
Shahid Raza\inst{2}\orcidlink{0000-0001-8192-0893}}

\authorrunning{J. Ménétrey et al.}

\institute{
University of Neuchâtel, Neuchâtel, Switzerland,
\email{first.last@unine.ch} \and
RISE Research Institutes of Sweden, Stockholm, Sweden,
\email{first.last@ri.se}
}

\maketitle

\begin{abstract}
Attestation is a fundamental building block to establish trust over software systems.
When used in conjunction with trusted execution environments, it guarantees the genuineness of the code executed against powerful attackers and threats, paving the way for adoption in several sensitive application domains.
This paper reviews remote attestation principles and explains how the modern and industrially well-established trusted execution environments Intel SGX, Arm TrustZone and AMD SEV, as well as emerging RISC-V solutions, leverage these mechanisms.

\end{abstract}

\sloppy

\def\confname{22nd International Conference on Distributed Applications and Interoperable Systems (DAIS'22)}
\def\confyear{2022}
\def\confdoi{XXX}

\definecolor{yellowPaper}{HTML}{fff8ae}
\AddToShipoutPictureFG*{\AtTextUpperLeft{\adjustbox{raise=59pt,center}{
    \begin{tcolorbox}[width=1.5\textwidth,colback=yellowPaper,enhanced,frame hidden,sharp corners]  
        \centering\scriptsize
        \copyright~\confyear\ Springer. Personal use of this material is permitted.
        Permission from Springer must be obtained for all other uses, in any current or future media, including reprinting/republishing this material for advertising or promotional purposes, creating new collective works, for resale or redistribution to servers or lists, or reuse of any copyrighted component of this work in other works.
        This is the author's version of the work.
        The final authenticated version is available online at \href{https://doi.org/10.1007/978-3-031-16092-9_7}{doi.org/10.1007/978-3-031-16092-9\_7} and has been published in the proceedings of the 
        \confname.
     \end{tcolorbox}} 
  }}

\hypersetup{
    pdfcopyright={\copyright~\confyear\ Springer. Personal use of this material is permitted. Permission from Springer must be obtained for all other uses, in any current or future media, including reprinting/republishing this material for advertising or promotional purposes, creating new collective works, for resale or redistribution to servers or lists, or reuse of any copyrighted component of this work in other works.
    This is the author's version of the work. The final authenticated version is available online at https://doi.org/10.1007/978-3-031-16092-9_7 and has been published in the proceedings of the 
    \confname.}
}
 \section{Introduction}
Confidentiality and integrity are essential features when building secure computer systems.
This is particularly important when the underlying system cannot be fully trusted or controlled.
For example, video broadcasting software can be tampered with by end-users to circumvent digital rights management, or virtual machines are candidly open to the indiscretion of their cloud-based untrusted hosts.
The introduction of Trusted Execution Environments (TEEs), such as Intel SGX, AMD SEV, RISC-V and Arm TrustZone-A/-M, into commodity processors, significantly mitigates the attack surface against powerful attackers.
In a nutshell, TEEs let a piece of software be executed with stronger security guarantees, including privacy and integrity properties, without relying on a trustworthy operating system.
Each of these enabling technologies offers different degrees of guarantees that can be leveraged to increase the confidentiality and integrity of applications.

Remote attestation allows establishing a trusting relationship with a specific software by verifying its authenticity and integrity.
Through remote attestation, one ensures to be communicating with a specific, trusted (attested) program remotely.
TEEs can support and strengthen the attestation process, ensuring that programs are shielded against many powerful attacks by isolating critical security software, assets and private information from the rest of the system.
However, to the best of our knowledge, there is not a clear systematisation of attestation mechanisms supported by modern and industrially well-established TEEs.
Hence, the main contribution of this work is to describe the \sotaAdj best practices regarding remote attestation mechanisms of TEEs, covering a necessarily incomplete selection of TEEs, which includes the four major technologies available for commodity hardware, which are Intel SGX, Arm TrustZone-A/-M, AMD SEV and many emerging TEEs using the open ISA RISC-V.
We complement previous work~\cite{maene2017attestation,systexsota} with an updated analysis of TEEs (\eg introduction of Intel SGX and Arm TrustZone variations), a thorough analysis of remote attestation mechanisms and coverage of the upcoming TEEs of Intel and Arm.

 \section{Attestation}
\label{sec:att}

\subsection{Local attestation}
Local attestation enables a trusted environment to prove its identity to any other trusted environments hosted on the same system, respectively, on the same CPU if the secret provisioned for the attestation is bound to the processor.
The target environment that receives the local attestation request can assess whether the issued proof is genuine by verifying its authentication, usually based on a symmetric-key scheme, using a \emph{message authentication code} (MAC).
This mechanism is required to establish secure communication channels between trusted environments, often used to delegate computing tasks securely.
As an example, Intel SGX's remote attestation (detailed in Section \ref{sec:sgx}) leverages the local attestation to sign proofs in another trusted environment through a secure communication channel.

\subsection{Remote attestation}
\label{sec:remote-att}
Remote attestation allows to establish trust between different devices and provides cryptographic proofs that the executing software is genuine and untampered~\cite{coker2011principles}.
In the remainder, we adopt the terminology proposed by the IETF to describe remote attestation and related architectures~\cite{ietf-rats-architecture-12}.
Under these terms, \emph{a relying party} wishes to establish a trusted relationship with an \emph{attester}, thanks to the help of a \emph{verifier}.
The attester provides the state of its system, indicating the hardware and the software stack that runs on its device by collecting a set of \emph{claims} of trustworthiness.
A claim is a piece of asserted information collected by an attesting environment, \eg a TEE.
An example of claims is the code \emph{measurement}, (\ie, a cryptographic hash of the application's code) of an executing program within a TEE.
TEEs also create additional claims that identify the \emph{trusted computing base} (TCB is the amount of hardware and software that needs to be trusted), so verifiers are able to evaluate the genuineness of the platform.
Claims are collected and cryptographically signed to form \emph{evidence}, later observed and accepted (or denied) by the verifier.
Once the attester is proven genuine, the relying party can safely interact with it and transfer confidential data or delegate computations.

The problem of remotely attesting software has been extensively studied in academia, and industrial implementations already exist.
Three leading families of remote attestation methods exist: (i) software-based, (ii) hardware-based, and (iii) hybrid (software- and hardware-based).
Software-based remote attestation\extcite[seshadri2005pioneer]{seshadri2005pioneer,10.1007/978-3-540-74477-1_97,steiner2019towards} does not depend on any particular hardware.
This method is particularly adapted to low-cost use cases.
Hardware-based remote attestation relies on a \emph{root of trust}, which is one or many cryptographic values rooted in hardware to ensure that the claims are trustworthy.
Typically, a root of trust can be implemented using tamper-resistant hardware, such as a \emph{trusted platform module} (TPM)~\cite{6104065}, a \emph{physical unclonable function} (PUF) that prevents impersonations by using unique hardware marks produced at manufacture\extcite[6881436]{6881436,feng2018aaot}, or a hardware secret fused in a die (\eg CPU) exposed exclusively to the trusted environment.
Hybrid solutions combine hardware devices and software implementations\extcite[10.1145/3460120.3484532]{6800458,carpent2018remote,10.1145/3460120.3484532}, in an attempt to leverage advantages from both sides.
Researchers used hardware/software co-design techniques to propose a hybrid design with a formal proof of correctness~\cite{236230}.
Finally, remote attestation mechanisms are popular among the TEEs due to their carefully controlled environments and their ability to generate code measurements.
Section \ref{sec:atttee} delivers extensive analysis of the \sotaNoun of the TEEs, including their support for remote attestation.

\subsection{Mutual attestation}
Trusted applications may need stronger trust assurances by ensuring both ends of a secure channel are attested.
For example, when retrieving confidential data from a sensing IoT device (where data is particularly sensitive), the device must authenticate the remote party, while the latter must ensure the sensing device has not been spoofed or tampered with.
Mutual attestation protocols have been designed to appraise the trustworthiness of both end devices involved in a communication.
We also report how mutual attestation has also been studied in the context of TEEs~\cite{10.1145/3342559.3365334}, as we further detail in Section \ref{sec:atttee}.

\begin{table*}[!htb]
    \scriptsize
    \centering
    \setlength{\tabcolsep}{0pt}
    \rowcolors{1}{gray!10}{gray!0}
    \begin{tabularx}{\textwidth}{X|q{8mm}q{8mm}|q{8mm}q{8mm}|q{5mm}q{5mm}q{5mm}|q{11mm}q{11mm}q{11mm}q{11mm}}
        \toprule
        \rowcolor{gray!25}
        & \multicolumn{2}{c|}{\textbf{SGX}} & \multicolumn{2}{c|}{\textbf{TrustZone}} & \multicolumn{3}{c|}{\textbf{SEV}} & \multicolumn{4}{c}{\textbf{RISC-V}} \\
        \rowcolor{gray!25}
        \multirow{-2}{*}{\makecell[l]{\textbf{Features}}} & \rotatebox{90}{Client SGX} & \rotatebox{90}{Scalable SGX} & \rotatebox{90}{TrustZone-A} & \rotatebox{90}{TrustZone-M} & \rotatebox{90}{Vanilla} & \rotatebox{90}{SEV-ES} & \rotatebox{90}{SEV-SNP} & \rotatebox{90}{Keystone} & \rotatebox{90}{Sanctum} & \rotatebox{90}{TIMBER-V} & \rotatebox{90}{LIRA-V} \\
        \midrule
        Integrity & \compfull & \comppart & \compnone & \compnone & \compnone & \compnone & \comppart & \compfull & \compnone & \compnone & \compnone \\
        Freshness & \compfull & \comppart & \compnone & \compnone & \compnone & \compnone & \comppart & \compfull & \compnone & \compnone & \compnone \\
        Encryption & \compfull & \compfull & \compnone & \compnone & \compfull & \compfull & \compfull & \compfull & \compnone & \compnone & \compnone \\
        Unlimited domains & \compfull & \compfull & \compnone & \compfull & \comppart & \compfull & \compfull & \compfull & \compfull & \compfull & \compnone \\
        Open source & \comppart & \comppart & \comppart & \comppart & \compnone & \compnone & \compnone & \compfull & \compfull & \compfull & \compnone \\
        Local attestation & \compfull & \compfull & \compnone & \compnone & \compnone & \compnone & \compnone & \compnone & \compfull & \compfull & \compnone \\
        Remote attestation & \compfull & \compfull & \comppart & \comppart & \compfull & \compfull & \compfull & \compfull & \compfull & \compfull & \compfull  \\
        API for attestation & \compfull & \compfull & \comppart & \comppart & \compnone & \compnone & \compfull & \compfull & \compfull & \compfull & \compfull \\
        Mutual attestation & \compnone & \compnone & \comppart & \compnone & \compnone & \compnone & \compnone & \compnone & \compnone & \compnone & \compfull \\
        User-mode support & \compfull & \compfull & \compfull & \compfull & \compfull & \compfull & \compfull & \compfull & \compfull & \compfull & \compnone \\
        Industrial TEE & \compfull & \compfull & \compfull & \compfull & \compfull & \compfull & \compfull & \compnone & \compnone & \compnone & \compnone \\
        \rowcolor{gray!0}
        &&&&&&&&&&& \\
        \rowcolor{gray!0}
        &&&&&&&&&&& \\
        \rowcolor{gray!0}
        \multirow{-3}*{\makecell[l]{Isolation and\\attestation granularity}} & \multicolumn{2}{c|}{\multirow{-3}*{\makecell[c]{Intra-\\address\\space}}} & \multicolumn{2}{c|}{\multirow{-3}*{\makecell[c]{Secure\\world}}} & \multicolumn{3}{c|}{\multirow{-3}*{\makecell[c]{VM}}} & \multirow{-3}*{\makecell[c]{Secure\\world}} & \multicolumn{3}{c}{\multirow{-3}*{\makecell[c]{Intra-address space}}} \\
        \rowcolor{gray!10}
        &&&&&&&&&&& \\
        \rowcolor{gray!10}
        \multirow{-2}*{\makecell[l]{System support for\\isolation}} & \multicolumn{2}{c|}{\multirow{-2}*{\makecell[c]{$\mu$code +\\XuCode}}} & \multirow{-2}*{\makecell[c]{SMC}} & \multirow{-2}*{\makecell[c]{MPU}} & \multicolumn{3}{c|}{\multirow{-2}*{\makecell[c]{Firmware}}} & \multicolumn{2}{c}{\multirow{-2}*{\makecell[c]{SMC + PMP}}} & \multirow{-2}*{\makecell[c]{Tag +\\MPU}} & \multirow{-2}*{\makecell[c]{PMP}}\\
        \bottomrule
    \end{tabularx}
    \caption{\label{tab:features-comparison}Comparison of the \sotaAdj TEEs.}
    \vspace{-15pt}
    \bigskip
    \scriptsize
    \centering
\rowcolors{1}{gray!10}{gray!0}
    \begin{tabularx}{\textwidth}{m{87pt}X}
        \toprule
        \rowcolor{gray!25}
        \textbf{Feature} & \textbf{Description}\\
        \midrule
        Integrity & An active mechanism preventing DRAM of TEE instances from being tampered with. Partial fulfilment means no protection against physical attacks.\\
        Freshness & Protecting DRAM of TEE instances against replay and rollback attacks. Partial fulfilment means no protection against physical attacks.\\
        Encryption & DRAM of TEE instances is encrypted to assure that no unauthorised access or memory snooping of the enclave occurs.\\
        Unlimited domains & Many TEE instances can run concurrently, while the TEE boundaries (\eg isolation, integrity) between these instances are guaranteed by hardware. Partial fulfilment means that the number of domains is capped.\\
        Open source & Indicate whether the solution is either partially or fully publicly available.\\
        Local attestation & A TEE instance attests running on the same system to another instance.\\
        Remote attestation & A TEE instance attests genuineness to remote parties. Partial fulfilment means no built-in support but is extended by the literature.\\
        API for attestation & An API is available by the trusted applications to interact with the process of remote attestation. Partial fulfilment means no built-in support but is extended by the literature.\\
        Mutual attestation & The identity of the attester and the verifier are authenticated upon remote attestations. Partial fulfilment means no built-in support but is extended by the literature.\\
        User mode support & State whether the trusted applications are hosted in user mode, according to the processor architecture.\\
        Industrial TEE & Contrast the TEEs used in production and made by the industry from the research prototypes designed by the academia.\\
        \makecell[l]{Isolation and\\attestation granularity} & The level of granularity where the TEE operates for providing isolation and attestation of the trusted software.\\
        \makecell[l]{System support for \\isolation} & The hardware mechanisms used to isolate trusted applications.\\
        \bottomrule
    \end{tabularx}
    \caption{\label{tab:features-list}Features of the \sotaAdj TEEs.}
    \vspace{-11pt}
\end{table*}

\section{Issuing attestations using TEEs}
\label{sec:atttee}

Several solutions exist to implement hardware support for trusted computing, and TEEs are particularly promising.
Typically, a TEE consists of isolating critical components of the system, (\eg, portions of the memory), denying access to more privileged but untrusted systems, such as kernel and machine modes.
Depending on the implementation, it guarantees the confidentiality and the integrity of the code and data of trusted applications, thanks to the assistance of CPU security features.
This work surveys modern and prevailing TEEs from processor designers and vendors with remote attestation capabilities for commodity or server-grade processors, namely Intel SGX~\cite{cryptoeprint:2016:086}, AMD SEV~\cite{amd-sev}, and Arm TrustZone~\cite{pinto2019demystifying}.
Besides, RISC-V, an open ISA with multiple open-source core implementations, ratified the \emph{physical memory protection} (PMP) instructions, offering similar capabilities to memory protection offered by aforementioned technologies\extcite{riscvratifications}.
As such, we also included many emerging academic and proprietary frameworks that capitalise on standard RISC-V primitives, which are Keystone~\cite{10.1145/3342195.3387532}, Sanctum~\cite{197162}, TIMBER-V~\cite{Weiser2019TIMBERVTM} and LIRA-V~\cite{9474324}.
Finally, among the many other technologies in the literature, we omitted the TEEs lacking remote attestation mechanisms (\eg IBM PEF~\cite{10.1145/3447786.3456243}) as well as the TEEs not supported on currently available CPUs (\eg Intel TDX~\cite{tdx}, Realm~\cite{armrealm} from Arm CCA~\cite{armccatech}).

\pagebreak
\subsection{TEE cornerstone features}\vspace{-3pt}
We propose a series of cornerstone features of TEEs and remote attestation capabilities and compare many emerging and well-established \sotaAdj solutions in Table \ref{tab:features-comparison}.
Each feature is detailed in Table \ref{tab:features-list} and can either be missing (\compnone), partially (\comppart) or fully (\compfull) available.
Besides, we elaborate further on each TEE in the remainder of the section.

\subsection{Trusted environments and remote attestation}

The attestation of software and hardware components requires an environment to issue evidence securely.
This role is usually assigned to some software or hardware mechanism that cannot be tampered with.
These environments rely on the code measurement of the executed software and combine that claim with cryptographic values derived from the root of trust.
We analysed today's practices for the leading processor vendors for issuing cryptographically signed evidence.

\begin{figure}[t]
    \centering
    \includegraphics[scale=0.7]{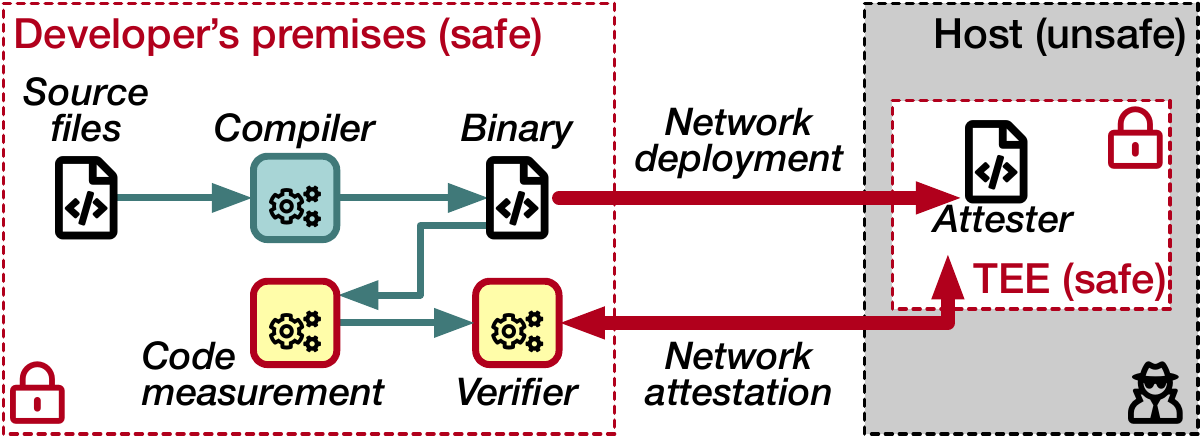}
    \caption{The workflow of deployment and attestation of TEEs.}
    \label{fig:workflow}
\end{figure}

Figure \ref{fig:workflow} illustrates the generic workflow TEE developers usually follow for the deployment of trusted applications.
Initially, the application is compiled and measured on the developers' premises.
It is later transferred to an untrusted system, executed in the TEE facility.
Once the trusted application is loaded and required to receive sensitive data, it communicates with a verifier to establish a trusted channel.
The TEE environment must facilitate this transaction by exposing evidence to the trusted application, which adds key material to bootstrap a secure channel from the TEE, thus preventing an attacker from eavesdropping on the communication.
The verifier examines the evidence, maintaining a list of reference values to identify genuine instances of trusted applications.
If recognised as trustworthy, the verifier can proceed to data exchanges.

\subsection{Intel SGX}
\label{sec:sgx}

\begin{figure*}[!t]
    \begin{subfigure}[b]{0.49\textwidth}
        \centering
        \includegraphics[width=\textwidth]{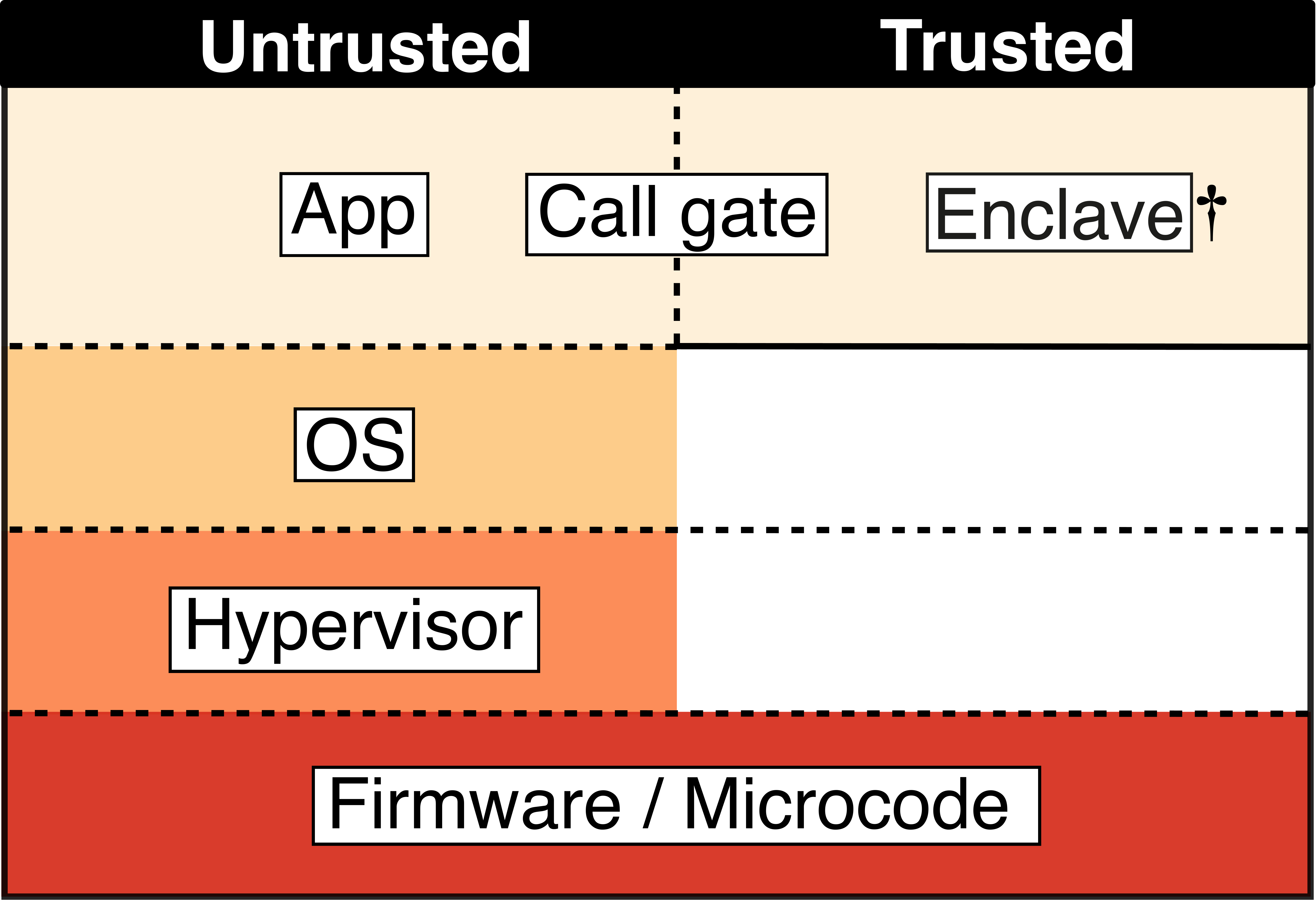}
        \caption{Intel SGX}
        \label{fig:sgx}
        \vspace{10pt}
    \end{subfigure}
    \hfill
    \begin{subfigure}[b]{0.49\textwidth}
        \centering
        \includegraphics[width=\textwidth]{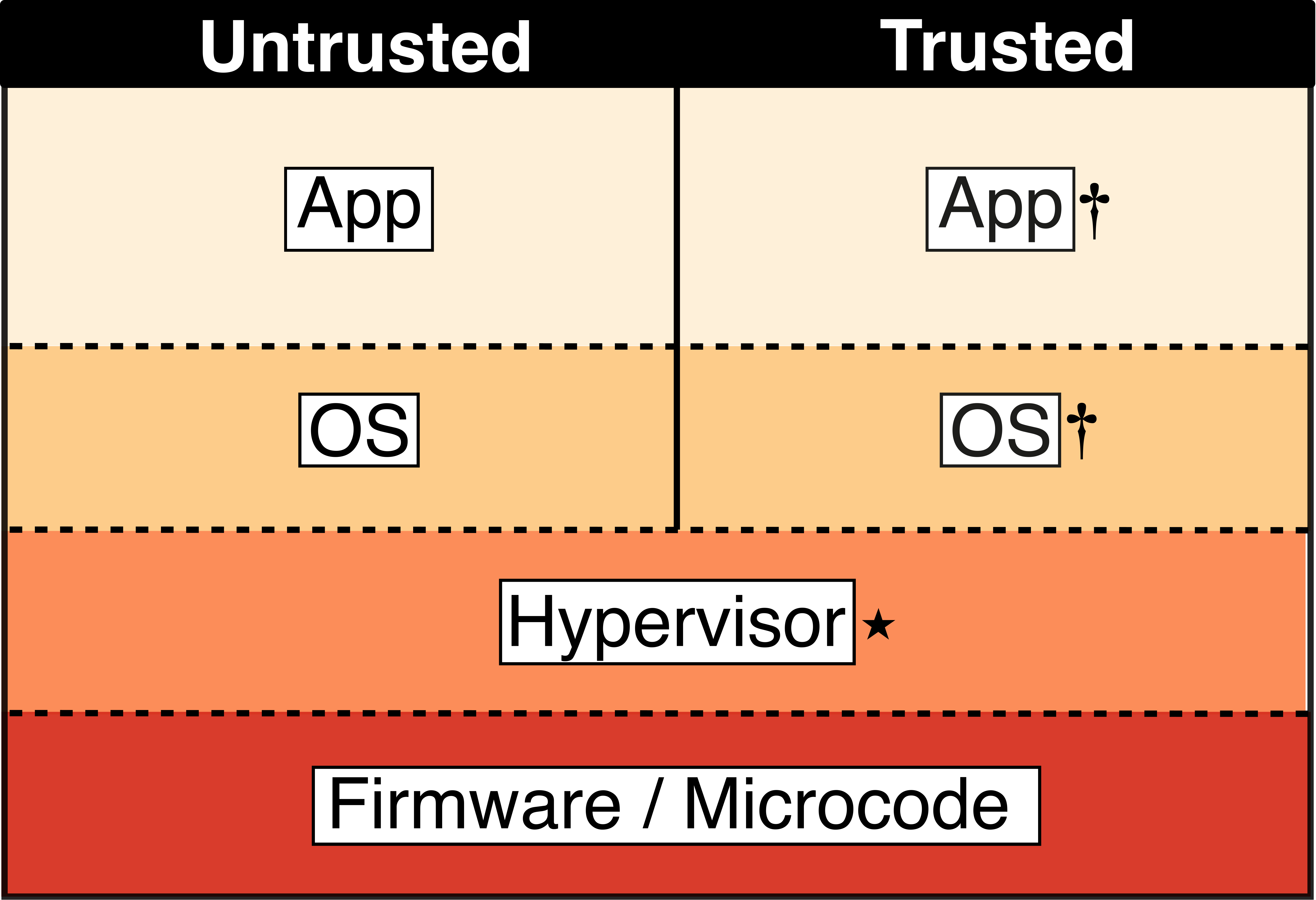}
        \caption{AMD SEV}
        \label{fig:sev}
        \vspace{10pt}
    \end{subfigure}
    \begin{subfigure}[b]{0.49\textwidth}
        \centering
        \includegraphics[width=\textwidth]{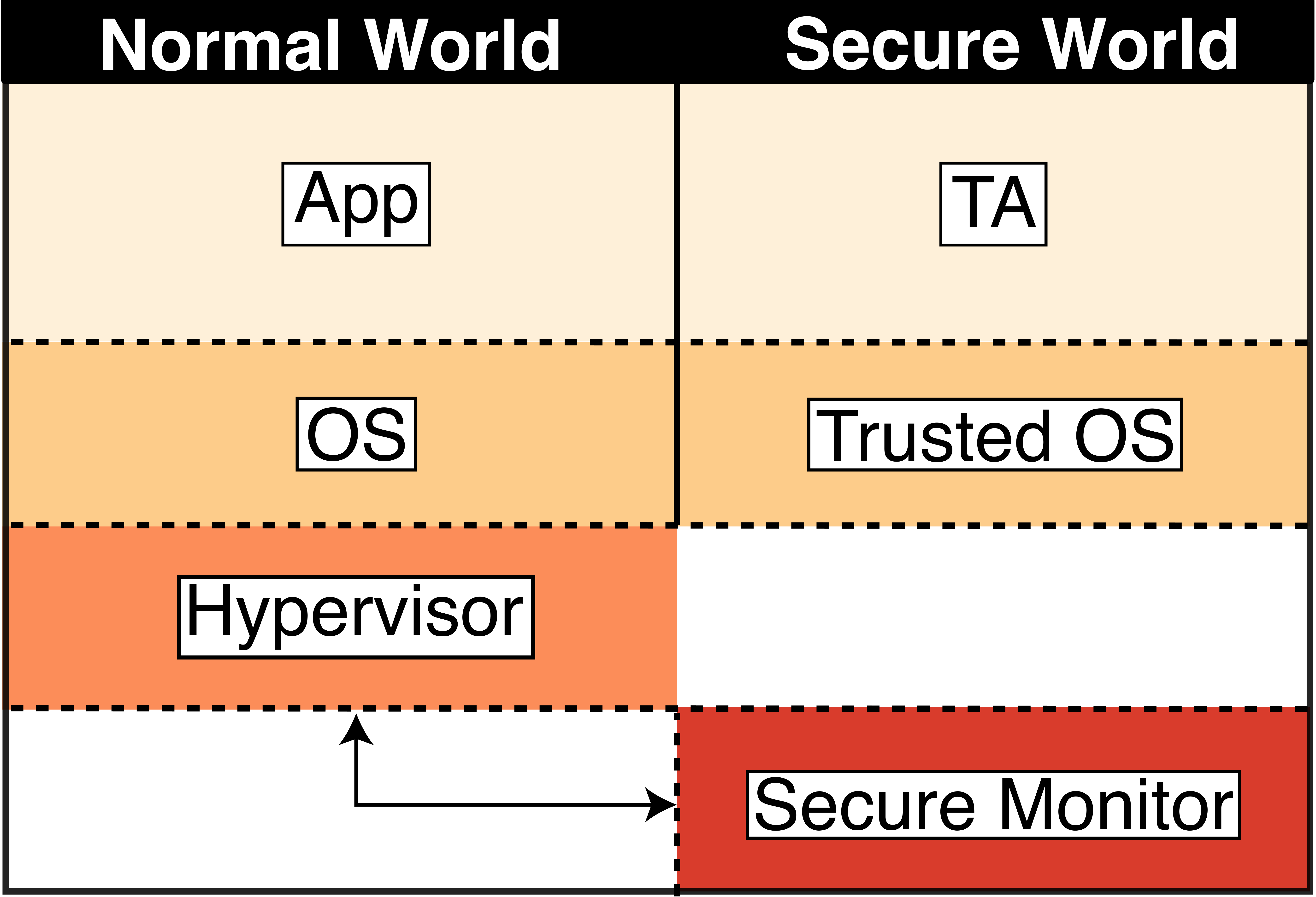}
        \caption{Arm TrustZone-A}
        \label{fig:trustzone-a}
    \end{subfigure}
    \hfill
    \begin{subfigure}[b]{0.49\textwidth}
        \centering
        \includegraphics[width=\textwidth]{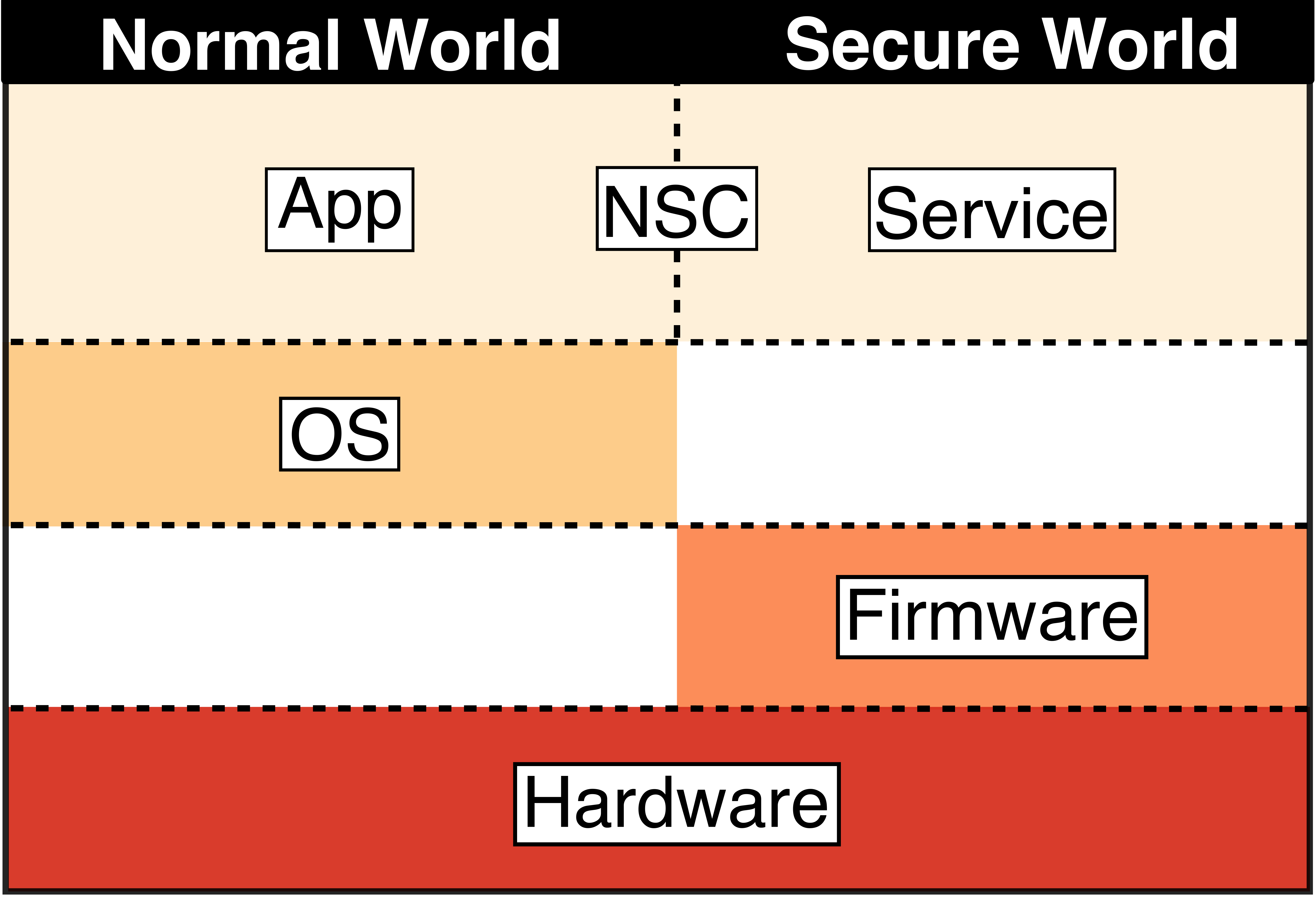}
        \caption{Arm TrustZone-M}
        \label{fig:trustzone-m}
    \end{subfigure}
    
    \vspace{6pt}
    \caption{Overview of the industrial TEE architectures.\\
    (\textdagger\ denotes the attested elements)\\
    (\textborn\ means trusted for SEV/SEV-ES, untrusted for SEV-SNP)}
    \label{fig:tees}
\end{figure*}

Intel Software Guard Extensions (SGX)~\cite{cryptoeprint:2016:086} introduced TEEs for mass-market processors in 2015.
Figure \ref{fig:sgx} illustrates the high-level architecture of SGX.
Specifically, Intel's Skylake architecture introduced a new set of processor instructions to create encrypted regions of memory, called \emph{enclaves}, living within the processes of the user space.
Intel SGX exist in two flavours: \emph{client} SGX and \emph{scalable} SGX~\cite{aublin2022scalableSGX}.
The former is the technology released in 2015, designed and implemented into consumer-grade processors, while the latter was released in 2021, focusing on server-grade processors.
The key differences between the two variants are: (i) the volatile memory available to enclaves, 128\,MB and 512\,GB, respectively, (ii) the multi-socket support and (iii) the lack of integrity and replay protections against hardware attacks for the latter.
Researchers conduct work to bring integrity protection for scalable SGX~\cite{aublin2022scalableSGX}.

These instructions are their own ISA that is implemented in XuCode~\cite{intel2021xucode} and together with \emph{model specific registers} provide the requirements to form the implementation of SGX.
XuCode is a technology that Intel developed and integrated into selected processor families to deliver new features more quickly and, particularly for SGX, reduce the impact a (complex) hardware implementation would have had on the features.
It operates from protected system memory in a special execution mode of the CPU, which are both set up by system firmware.
SGX is, to date, the only technology that is making use of XuCode.

A memory region is reserved at boot time for storing code and data of encrypted enclaves.
This memory area, called the \emph{enclave page cache} (EPC), is inaccessible to other programs running on the same machine, including the operating system and the hypervisor.
The traffic between the CPU and the system memory remains confidential thanks to the \emph{memory encryption engine} (MEE).
The EPC also stores verification codes to ensure that the DRAM corresponding to the EPC was not modified by any software external to the enclave.

A trusted application executing in an enclave may establish a local attestation with another enclave running on the same hardware.
Toward this end, Intel SGX issues a set of claims, called \emph{report}, that contains identities, attributes (\ie modes and other properties), the trustworthiness of the TCB, additional information for the target enclave and a MAC.
Unlike local attestation, remote attestation uses an asymmetric-key scheme, which is made possible by a special enclave, called \emph{quoting enclave}, that has access to the device-specific private key.
Intel designed their remote attestation protocol based on the SIGMA protocol~\cite{10.1007/978-3-540-45146-4_24} and extended it to the \emph{enhanced privacy ID} (EPID).
The EPID scheme does not identify unique entities, but rather a group of attesters.
Each attester belongs to a group, and the verifier checks the group's public key.
Evidence is signed by the EPID key, which guarantees the trustworthiness of the hardware and is bound to the firmware version of the processor.

\begin{figure}[!t]
    \centering
    \includegraphics[scale=0.2]{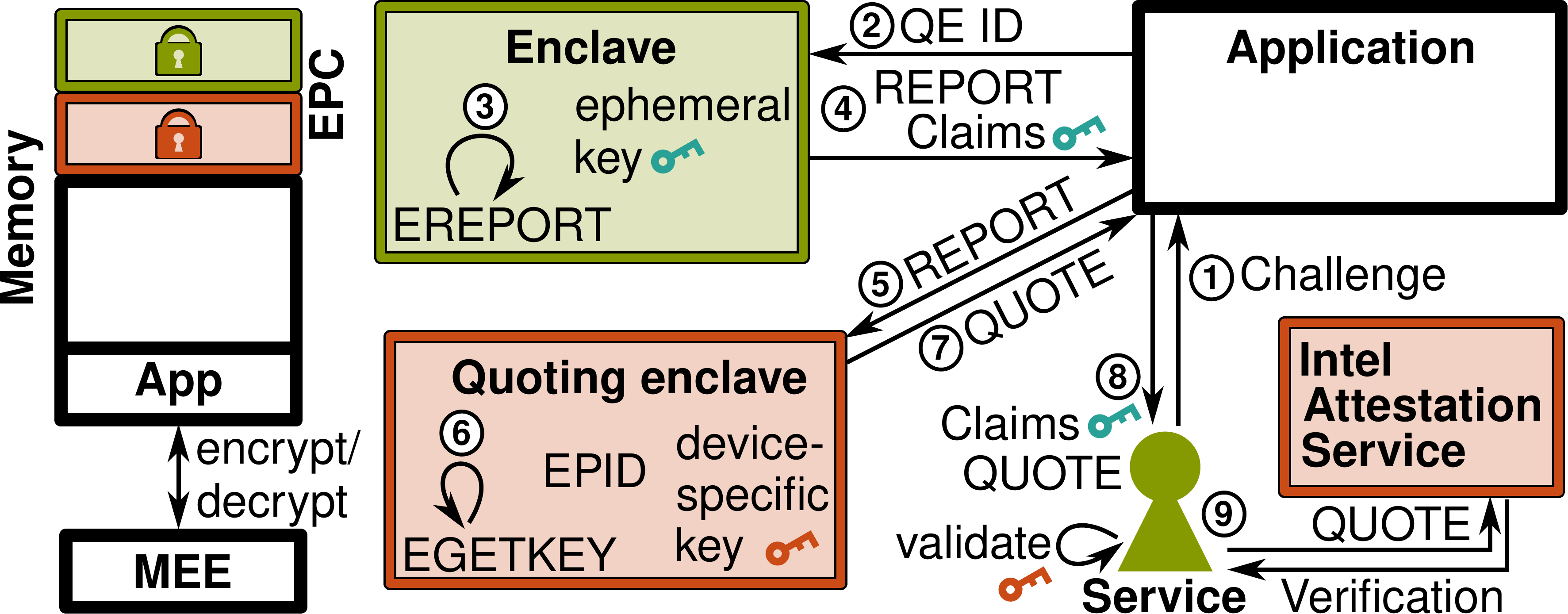}
    \caption{The remote attestation flow of Intel SGX.\label{fig:sgxra}}
\end{figure}

In a remote attestation scenario, a verifier submits a challenge to the attester (\ie application enclave) with a nonce (Fig.\labelcref{fig:sgxra}-\ding{192}).
The attester prepares a response to the challenge by creating a set of claims, a public key (Fig.\labelcref{fig:sgxra}-\ding{194}), and performs a local attestation with the quoting enclave.
After verifying the set of claims (\ie report), the quoting enclave signs the report to form evidence with the EPID key obtained using the \texttt{EGETKEY} instruction (Fig.\labelcref{fig:sgxra}-\ding{197}) and returns the evidence to the attester (Fig.\labelcref{fig:sgxra}-\ding{198}), which sends it back to the verifier (Fig.\labelcref{fig:sgxra}-\ding{199}).
The public key contained in the evidence enables the creation of a confidential communication channel.
Finally, the verifier examines the signature of the evidence (Fig.\labelcref{fig:sgxra}-\ding{200}) using the Intel attestation service (IAS)~\cite{anati2013innovative,brickell2007enhanced}.
If deemed trustworthy, the verifier may provision sensitive data to the attester using the secure channel.

More recently, Intel introduced the Data Center Attestation Primitives (DCAP)~\cite{scarlata2018supporting}, an alternative solution to EPID, enabling third-party attestation.
Thanks to DCAP, the verifiers have their own attestation infrastructure and prevent depending on external dependencies (\eg IAS) during the attestation procedure.
DCAP introduces an additional step, where the quote (Fig.\labelcref{fig:sgxra}-\ding{198}) is signed using \emph{elliptic curve digital signature algorithm} (ECDSA) by the attestation collateral of the attestation infrastructure.
Instead of contacting the IAS (Fig.\labelcref{fig:sgxra}-\ding{200}), the service retrieves the attestation collateral associated with the received evidence from the attestation infrastructure in order to validate the signature.

While the quoting enclave, the microcode and XuCode are closed-source, recent work analysed the TEE and its attestation mechanism formally~\cite{10.1145/3133956.3134098,9217791}.
The other components of SGX (\ie kernel driver and SDK) are open source.
MAGE~\cite{chen2020mage} further extended the remote attestation scheme of Intel SGX by offering mutual attestation for a group of enclaves without trusted third parties.
Similarly, OPERA~\cite{10.1145/3319535.3354220} proposes a decentralised attestation scheme, unchaining the attesters from the IAS while conducting attestation.

Intel SGX has many advantages but suffers from a few limitations as well.
First, most of the SGX implementation limits the EPC size to 93.5\,MB~\cite{vaucher2018sgx}.
While smaller programs offer smaller attack surfaces, exceeding this threshold increases the memory access latency because of its pagination mechanism.
Newer Intel Xeon processors extend that limit to 512\,GB, but drop integrity protection and freshness against physical attacks.
Besides, the enclave model prevents performing system calls and direct hardware access since the threat model distrusts the outer world, leading to the development of partitioned applications.

\subsection{Arm TrustZone architectures}
Depending on the architecture of Arm's processors, TrustZone comes in two flavours: TrustZone-A (for Cortex-A) and TrustZone-M (for Cortex-M).
While they share many design aspects, we detail how different they are in the remainder.

\subsubsection{Arm TrustZone-A} provides the hardware elements to establish a single TEE per system~\cite{pinto2019demystifying}.
Figure \ref{fig:trustzone-a} illustrates the high-level architecture of TrustZone-A.
Broadly adopted by commodity devices, TrustZone splits the processor into two states: the secure world (TEE) and the normal world (untrusted environment). A \emph{secure monitor} (SMC) is switching between worlds, and each world operates with its own user and kernel spaces.
The trusted world uses a trusted operating system (\eg, OP-TEE\extcite{optee}) and runs \emph{trusted applications} (TAs) as isolated processes.
The normal world uses a traditional operating system (\eg, Linux).

Despite the commercial success of TrustZone-A, it lacks attestation mechanisms, preventing relying parties from validating and trusting the state of TrustZone-A remotely.
Nevertheless, researchers proposed several variants of one-way remote attestation protocols for Arm TrustZone~\cite{10.1145/3319535.3363205,10.1145/2742647.2742676}, as well as mutual remote attestation~\cite{ahn2020design,10.1145/3098954.3098971}, thus extending the built-in capabilities of the architecture for attestation.
All of these propositions require the availability of hardware primitives on the \emph{system-on-chip} (SoC):
(i)~a root of trust in the secure world,
(ii)~a secure source of randomness for cryptographic operations, and
(iii)~a secure boot mechanism, ensuring the sane state of a system upon boot.
Indeed, devices lacking built-in attestation mechanisms may rely on a root of trust to derive private cryptographic materials (\eg a private key for evidence issuance).
Secure boot measures the integrity of individual boot stages on devices and prevents tampered systems from being booted.
As a result, remote parties can verify issued evidence in the TEE and ensure the trustworthiness of the attesters.

We describe the remote attestation mechanism of Shepherd et al.~\cite{10.1145/3098954.3098971} as a study case.
This solution establishes mutually trusted channels for bi-directional attestation, based on a \emph{trusted measurer} (TM), which is a software component located in the trusted world and authenticated by the TEE's secure boot, to generate claims and issue evidence based on the OS and TA states.
A private key is provisioned and sealed in the TEE's secure storage and used by the TM to sign evidence, similarly to a firmware TPM~\cite{197213}.
Using a dedicated protocol for remote attestation, the bi-directional attestation is accomplished in three rounds:
\begin{enumerate}
    \item The attester sends a handshake request to the verifier containing the identity of both parties and the cryptographic materials to initiate keys establishment.
    \item The verifier answers to the handshake by including similar information (\ie both identifies and cryptographic materials), as well as evidence of the verifier's TEE, based on the computed common secret (\ie using Diffie-Hellman).
    \item Finally, the attester sends back signed evidence of the attester's TEE, based on the same common secret. 
\end{enumerate}
Once the two parties validated the genuineness of the evidence, they can derive further shared secrets to establish a trusted communication channel.

Arm TrustZone-A also presents some advantages and drawbacks.
Hardware is independently accessible by both worlds, which is helpful for TEE applications utilising peripherals.
On the other hand, the reference and open-source trusted OS, \ie OP-TEE, limits the memory available to TAs by a few MB~\cite{10.1007/978-3-030-22496-7_11}.
Due to this constraint, software needs to be partitioned to leverage TrustZone.
Besides, the system must be installed in a particular way: a trusted OS is required, instead of creating TEE instances directly in the regular OS, bringing more complexity.
Finally, OP-TEE is small and does not implement a POSIX API, making developing TAs difficult, notably when porting legacy code.
While most components of TrustZone have open-source alternatives (\eg the firmware and the trusted OS), many vendors do not disclose the implementation of the secure monitor. 
\subsubsection{Arm TrustZone-M}

(TZ-M) much like its predecessor TrustZone-A, provides an efficient mechanism to isolate the system into two distinct states/processing environments\extcite[armv8m]{armv8m,trustzoneform}.
The TZ-M extension brings trusted execution into resource-constrained IoT devices (\eg Cortex-M23/M33/M35P/M55).
When a TZ-M enabled device boots up, it always starts in the secure world, where the memory is initialised before transferring the control to the normal world.
Despite the similarity regarding the high-level concept, TrustZone-M differs from TrustZone-A in low-level implementation of some features.
The switch between the secure and the normal world is embedded in hardware and is much faster than the secure monitor\extcite[armv8march]{sg,armv8march}.
This makes the context switching efficient and suitable for constrained devices.
The normal world applications directly call the secure world functions using the \emph{non-secure callable} (NSC) region (Figure \ref{fig:trustzone-m}).
TrustZone-M lacks complex memory management operations like the \emph{memory management unit} (MMU) and only supports the \emph{memory protection unit} (MPU) to enforce even finer levels of access control and memory protection~\cite{armmpu}.
In TZ-M enabled IoT devices, the secure world runs a concise trusted firmware which provides secure processing in the form of secure services (\eg TrustedFirmware-M\extcite{tfm}), which is a reference implementation of Platform Security Architecture (PSA)~\cite{armpsa}) and the normal world supports real-time operating systems (\eg Zephyr OS\extcite{zephyr}, Arm MBED OS\extcite{mbed}, FreeRTOS\extcite{freertos}).

Since TZ-M is a relatively new addition, recently available for the IoT infrastructure, existing work on attestation mechanisms for the hardware/software is scarce.
Nonetheless, TZ-M fulfils some basic requirements for attestation like (i) secure storage, (ii) secure boot, (iii) secure inter-world communication and (iv) isolation of software.
Thus, schemes like~\cite{abera2019diat} have leveraged TZ-M to develop attestation and use TZ-M's TEE capabilities to establish a chain of trust.
TrustedFirmware-M, following the guidelines of PSA, also supports initial attestation of device-specific data in the form of a secure service\extcite{psaattest,attestveriferietf}.
We provide further details of the remote attestation mechanism introduced in DIAT~\cite{abera2019diat}.
It aims at providing run-time attestation of on-device data integrity in autonomous embedded systems in the absence of a central verifier.
They provide attestation of the data integrity by identifying the software components (or modules), \ie the claims, that process the data of concern, verifying that the modules are not modified, ensuring that all modules of software that influence data are benign.
Data integrity is provided by attestation, ensuring correct processing of the sensitive data.
The main steps of the protocol are described below:

\begin{itemize}
    \item The verifier sends a request for data to the attester along with a nonce. The data can represent collected environmental (\eg a sensing edge device) or compute-bound (\eg machine learning) information.
    \item The attester generates the requested data and issues evidence, called the \emph{attestation results}, which are the list of all the software modules that affect the data, and the control flow of each module is derived using the control flow graph.
 The attester signs the data and the evidence with its secret key and sends the authenticated data to the verifier.
    \item The verifier assesses the authenticity and integrity of the data by tracing the software modules from the evidence. Since the evidence is comprised of software modules that process the data and the frequency of execution of a module, unauthorised data modifications and code reuse attacks are detected and prevented.
\end{itemize}

TrustZone-M provides several advantages as a TEE to support remote attestation but also has a few drawbacks.
It provides efficient isolation of the software modules and a faster context switch between the secure and normal world.
This is advantageous as it is critical to have minimum attestation latency in the real-time operations of embedded systems like autonomous vehicles, industrial control systems, unmanned aerial vehicles, etc.
The availability of hardware-unique keys in TZ-M enabled devices further ensures that the evidence generated by the TCB cannot be forged.
Besides, the software stack may be fully open source, thanks to the absence of a secure monitor.
On the other hand, since the components involved in measuring, attesting, and verifying the data/system need to be protected as part of the TCB, it increases the TCB size on the attested devices, raising the attack surface.

\subsection{AMD SEV}

AMD Secure Encrypted Virtualization (SEV)~\cite{amd-sev} allows isolating virtualised environments (\eg containers and virtual machines) from trusted hypervisors.
Figure \ref{fig:sev} illustrates the high-level architecture of SEV.
SEV uses an embedded hardware AES engine, which relies on multiple keys to encrypt memory seamlessly. It exploits a closed Arm Cortex-v5 processor as a secure co-processor, used to generate cryptographic materials kept in the CPU.
Each virtual machine (VM) and hypervisor is assigned a particular key and tagged with an \emph{address space identifier} (ASID), preventing cross-TEE attacks.
The tag restricts the code and data usage to the owner with the same ASID and protects from unauthorised usage inside the processor.
Code and data are protected by AES encryption with a 128-bit key based on the tag outside the processor package.

The original version of SEV could leak sensitive information during interrupts from guests to the hypervisor through registers~\cite{hetzelt2017sev}.
This issue was addressed with SEV Encrypted State (SEV-ES)~\cite{kaplan2017seves}, where register states are encrypted, and the guest operating system needs to grant the hypervisor access to specific guest registers.
Register states are stored with SEV-ES for each VM in a \emph{virtual machine control block} (VMCB) that is divided into an unencrypted control area and an encrypted \emph{virtual machine save area}. The hypervisor manages the control area to indicate event and interrupt handling, while VMSA contains register states.
Integrity protection ensures that encrypted register values in the VMSA cannot be modified without being noticed and VMs resume with the same state.
Requesting services from the hypervisor due to interrupts in VMs are communicated over the \emph{guest hypervisor communication block} (GHCB) that is accessible through shared memory.
Hypervisors do not need to be trusted with SEV-ES because they no longer have access to guest registers.
However, the remote attestation protocol was recently proven unsecure~\cite{buhren2019insecure}, exposing the system to rollback attacks and allowing a malicious cloud provider with physical access to SEV machines to easily install malicious firmware and be able to read in clear the (otherwise protected) system.
Future iterations of this technology, \ie, SEV Secure Nested Paging (SEV-SNP)~\cite{sev2020strengthening}, plan to overcome these limitations, typically by means of in-silico redesigns.

At its core, SEV leverages a root of trust, called \emph{chip endorsement key}, a secret fused in the die of the processor and issued by AMD for its attestation mechanism.
The three editions of SEV may start the VMs from an unencrypted state, similarly to Intel SGX enclaves.
In such cases, the secrets and confidential data must then be provisioned using remote attestation.
The AMD secure processor creates a claim based on the measurement of the content of the VM.
In addition, SEV-SNP measures the metadata associated with memory pages, ensuring the digest also considers the layout of the initial guest memory.
While SEV and SEV-ES only support remote attestation during the launch of the guest operating system, SEV-SNP supports a more flexible model.
That latter bootstraps private communication keys, enabling the guest VM to request evidence at any time and obtain cryptographic materials for data sealing, \ie, storing data securely at rest.

\begin{figure}[!t]
  \centering
  \includegraphics[scale=0.2]{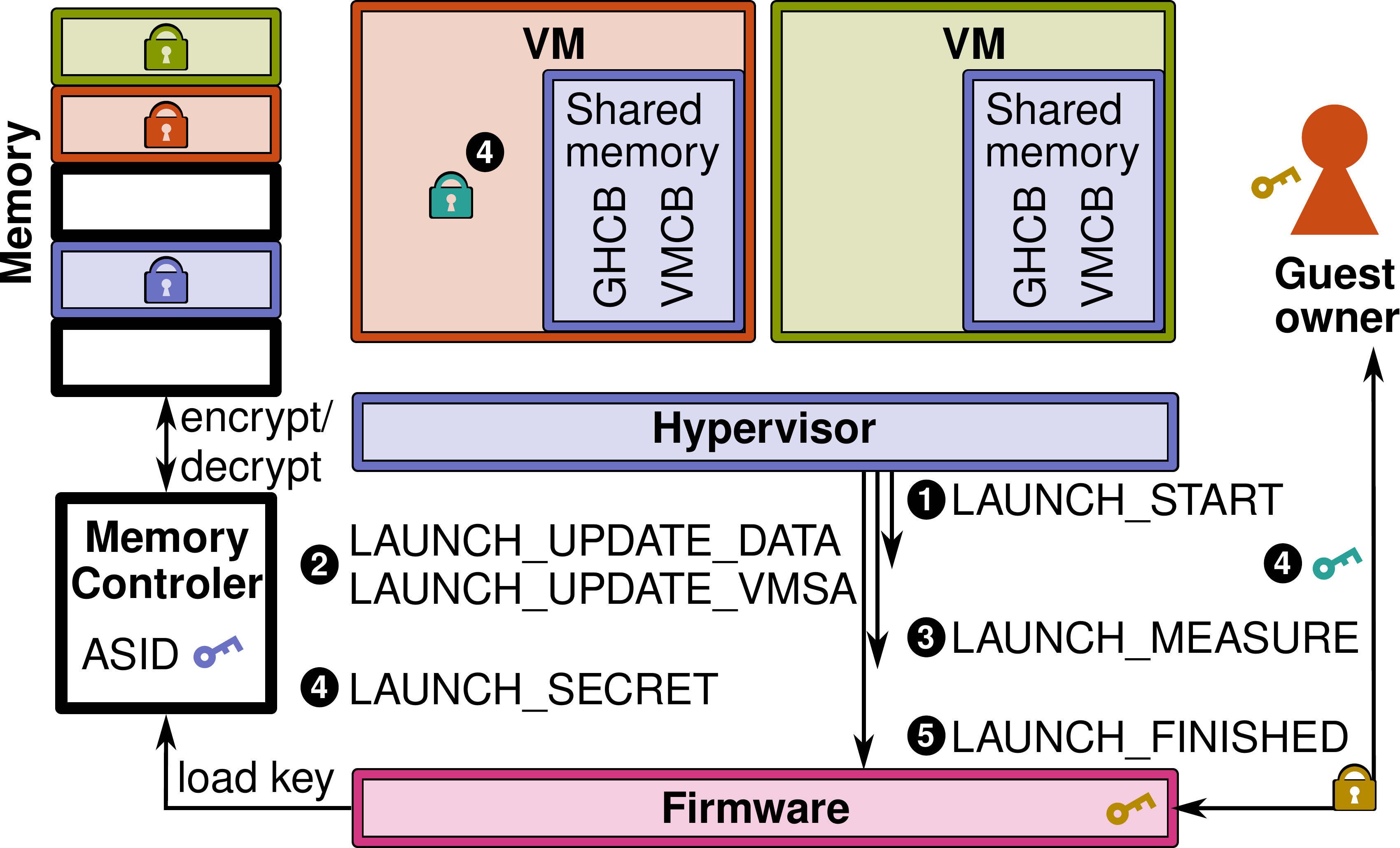}
  \caption{The remote attestation flow of AMD SEV.\label{fig:sevra}}
\end{figure}

The remote attestation process takes place when SEV is starting the VMs.
First, the attester, called \emph{hypervisor}, executes the \texttt{LAUNCH\_START} command (Fig.\labelcref{fig:sevra}-\ding{202}) which creates a guest context in the firmware with the public key of the verifier, called \emph{guest owner}.
As the attester is loading the VM into memory, the \texttt{LAUNCH\_UPDATE\_DATA}/\texttt{LAUNCH\_UPDATE\_VMSA} commands (Fig.\labelcref{fig:sevra}-\ding{203}) are called to encrypt the memory and calculate the claims.
When the VM is loaded, the attester calls the \texttt{LAUNCH\_MEASURE} command (Fig.\labelcref{fig:sevra}-\ding{204}), which produces evidence of the encrypted VM. The SEV firmware provides the verifier with evidence of the state of the VM to prove that it is in the expected state.
The verifier examines the evidence to determine whether the VM has not been interfered with.
Finally, sensitive data, such as image decryption keys, is provisioned through the \texttt{LAUNCH\_SECRET} command (Fig.\labelcref{fig:sevra}-\ding{205}) after which the attester calls the \texttt{LAUNCH\_FINISHED} command (Fig.\labelcref{fig:sevra}-\ding{206}) to indicate that the VM can be executed.

Software development is eased, as AMD SEV protects the whole VM, which comprises the operating system, unlike Intel SGX, where the applications are split into untrusted and trusted parts.
Nonetheless, this approach increases the attack surface of the secure environment since the TCB is enlarged.
The guest operating system must also support SEV, cannot access host devices (PCI passthrough), and the first edition of SEV (called \emph{vanilla} in Table \ref{tab:features-comparison}) is limited to 16 VMs.

\subsection{RISC-V architectures}
There exist several proposals for TEEs designs for RISC-V based on PMP instructions.
These proposals include support for remote attestation, such as those previously described. 
We survey the most important ones in the following.

Keystone~\cite{10.1145/3342195.3387532} is a modular framework that provides the building blocks to create trusted execution environments, rather than providing an all-in-one solution that is inflexible and is another fixed design point.
Instead, they advocate that hardware should provide security primitives instead of point-wise solutions.
Keystone implements a secure monitor at machine mode (M-mode) and relies on the RISC-V PMP instructions to provide isolated execution and, therefore, does not require any hardware change.
Since Keystone leverages features composition, the framework users can select their own set of security primitives, \eg, memory encryption, dynamic memory management and cache partitioning.
Each trusted application executes in user mode (U-mode) and embeds a runtime that executes in supervisor mode (S-mode).
The runtime decouples the infrastructure aspect of the TEE (\eg, memory management, scheduling) from the security aspect handled by the secure monitor.
As such, Keystone programmers can roll their custom runtime to fine-grained control of the computer resources without managing the TEE's security.
Keystone utilises a secure boot mechanism that measures the secure monitor image, generates an attestation key and signs them using a root of trust.
The secure monitor exposes a \emph{supervisor system interface} (SBI) for the enclaves to communicate.
A subset of the SBI is dedicated to issue evidence signed by provisioned keys (\ie endorsed by the verifier), based on the measurement of the secure monitor, the runtime and the enclave's application.
Arbitrary data can be attached to evidence, enabling an attester to create a secure communication channel with a verifier using key establishment protocols (\eg Diffie-Hellman).
When a remote attestation request takes place, the verifier sends a challenge to the trusted application.
The response contains evidence with the public session key of the attester.
Finally, the verifier examines the evidence based on the public signature and the claims (\ie measurements of components), leading to establishing a secure communication channel.
While Keystone does not describe in-depth the protocol, the authors provide a case study of remote attestation.

Sanctum~\cite{197162} has been the first proposition with support for attesting trusted applications.
It offers similar promises to Intel's SGX by providing provable and robust software isolation, running in enclaves.
The authors replaced Intel's opaque microcode/XuCode with two open-source components: the \emph{measurement root} (\texttt{mroot}) and a secure monitor to provide verifiable protection.
A remote attestation protocol is proposed, as well as a comprehensive design for deriving trust from a root of trust.
Upon booting the system, \texttt{mroot} generates the cryptographic materials for signing if started for the first time and hands off to the secure monitor.
Similarly to SGX, Sanctum utilises a \emph{signing enclave}, that receives a derived private key from the secure monitor for evidence generation.
The remote attestation protocol requires the attester, called \emph{enclave}, to establish a session key with a verifier, called \emph{remote party}.
Afterwards, an enclave can request evidence from the signing enclave based on multiple claims, such as the hash of the code of the requesting enclave and some information coming from the key exchange messages.
The evidence is then forwarded to the verifier through the secure channel for examination.
This work has been further extended to establish a secure boot mechanism and an alternative method for remote attestation by deriving a cryptographic identity from manufacturing variation using a PUF, which is useful when a hardware secret is not present~\cite{8429295}.

TIMBER-V~\cite{Weiser2019TIMBERVTM} achieved the isolation of execution on small embedded processors thanks to hardware-assisted memory tagging.
Tagged memory transparently associates blocks of memory with additional metadata.
Unlike Sanctum, they aim to bring enclaves to smaller RISC-V featuring only limited physical memory.
Similarly to TrustZone, the user mode (U-mode) and the supervisor mode (S-mode) are split into a secure and normal world.
The secure supervisor mode runs a trust manager, called \emph{TagRoot}, which manages the tagging of the memory.
The secure user mode improves the model of TrustZone, as it can handle multiple concurrent enclaves, which are isolated from each other.
They combine tagged memory with an MPU to support an arbitrary number of processes while avoiding the overhead of large tags.
The trust manager exposes an API for the enclaves to retrieve evidence, based on a given enclave identity, a root of trust, called the \emph{secret platform key}, and an arbitrary identifier provided by the enclave.
The remote attestation protocol is twofold: the verifier (\ie remote party) sends a challenge to the attester (\ie enclave).
Next, the challenge is forwarded to the trust manager as an identifier to issue evidence, which is authenticated using a MAC.
The usage of symmetric cryptography is unusual in remote attestation because the verifier requires to own the secret key to verify the evidence.
The authors added that TIMBER-V could be extended to leverage public-key cryptography for remote attestation.

LIRA-V~\cite{9474324} drafted a mutual remote attestation for constrained edge devices.
While this solution does not enable the execution of arbitrary code in a TEE, it introduces a comprehensive remote attestation mechanism that leverages PMP for code protection of the attesting environment and the availability of a root of trust to issue evidence.
The proposed protocol relies exclusively on machine mode (M-mode) or machine and user mode (M-mode and U-mode).
The claim, which is the code measurement, is computed on parts of the physical memory regions by a program stored in the ROM.
LIRA-V's mutual attestation protocol works similarly to the protocol illustrated in TrustZone-A, in three rounds and requires provisioned keys as a root of trust.
The first device (\ie verifier) sends a challenge with a public session key.
Next, the second device (\ie attester) answers with a challenge and public session key, as well as evidence bound to that device and encrypted using the established shared session key.
Finally, if the first device validates the evidence, it becomes the attester and issues evidence for the second device, which becomes the verifier.
This protocol has been formally verified and enables the creation of a trusted communication channel upon the validation of evidence.

Lastly, we omitted some other emerging TEEs leveraging RISC-V as they lack remote attestation mechanisms.
These technologies are yet to be researched for bringing such capabilities.
We briefly introduce them here for completeness.
SiFive, the provider of commercial RISC-V processor IP, proposes Hex-Five MultiZone~\cite{garlati2020clean}, a zero-trust computing architecture enabling the isolation of software, called \emph{zones}.
The multi zones kernel ensures the sane state of the system using secure boot and PMP and runs unmodified applications by trapping and emulating functionality for privileged instructions.
HECTOR-V~\cite{10.1145/3433210.3453112} is a design for developing hardened TEEs with a reduced TCB. Thanks to a tight coupling of the TEE and the SoC, the authors provide runtime and peripherals services directly from the hardware and leverage a dedicated processor and a hardware-based security monitor, which ensure the isolation and the control-flow integrity of the trusted applications, called \emph{trustlets}.
Finally, Lindemer et al.~\cite{lindemer2020real} enable simultaneous thread isolation and TEE separation on devices with a flat address space (\ie without an MMU), thanks to a minor change in the PMP specification.  \section{Future work}
TEEs and remote attestation are fast-moving research areas, where we expect many technological and paradigm enhancements in the next decades.
This section introduces the next trusted environments announced by Intel and Arm.
Besides, we also describe a shift to VM-based TEEs and conclude on attestation uniformity. 

Intel unveiled Trust Domain Extensions (TDX)~\cite{tdx} in 2020 as its upcoming TEE, introducing the deployment of hardware-isolated virtual machines, called \emph{trust domains}.
Similarly to AMD SEV, Intel TDX is designed to isolate legacy applications running on regular operating systems, unlike Intel SGX, which requires tailored software working on a split architecture (\ie untrusted and trusted parts).
TDX leverages Intel Virtual Machine Extensions and Intel Multi-Key Total Memory Encryption, as well as proposes an attestation process to guarantee the trustworthiness of the trust domains for relying parties.
In particular, it extends the remote attestation mechanisms of Intel SGX to issue claims and evidence, which has been formally verified by researchers~\cite{sardar2021demystifying}.

Arm announced Confidential Compute Architecture (CCA)~\cite{armccatech} as part of their future Armv9 processor architecture, consolidating TrustZone to isolate secure virtual machines.
With this aim in mind, Arm CCA leverages Arm Realm Management Extension~\cite{armrealm} to create a trusted third world called \emph{realm}, next to the existing normal and secure worlds.
Arm designed CCA to provide remote attestation mechanisms, assuring that relying parties can trust data and transactions.

These two recent initiatives highlight a convergence into the VM-based isolation paradigm.
Initially started by AMD, that architecture of TEEs has many advantages.
In particular, it reduces the developers' friction in writing applications, since the underlying operating system and API are standard and no different compared to the outside of the TEE.
Furthermore, a convergence of the paradigm may ease the development of unified and hardware-agnostic solutions for trusted software deployment, such as Open Enclave SDK~\cite{oesdk} or the recent initiatives promoting WebAssembly as an abstract portable executable code running in TEEs~\cite{enarx,veracruz,menetrey2021twine}.
Remote attestation may also benefit from these unified solutions by abstracting the attestation process behind standard interfaces.

 \section{Conclusion}
\label{sec:conc}
This work compares \sotaAdj remote attestation schemes, which leverage hardware-assisted TEEs, which help deploy and run trusted applications from commodity devices to cloud providers.
TEE-based remote attestation has not yet been extensively studied and remains an industrial challenge.

Our survey highlights four architectural extensions: Intel SGX, Arm TrustZone, AMD SEV, and upcoming RISC-V TEEs.
While SGX competes with SEV, the two pursue significantly different approaches.
The former provides a complete built-in remote attestation protocol for multiple, independent, trusted applications.
The latter is designed for virtualised environments, shielding VMs from untrusted hypervisors, and provides instructions to help the attestation of independent VMs.
Arm TrustZone and native RISC-V do not provide means for attesting software running in the trusted environment, relying on the community to develop alternatives.
However, TrustZone-M supports a root of trust, helping to develop an adequately trusted implementation.
RISC-V extensions differ a lot, offering different combinations of software and hardware extensions, some of which support a root of trust and multiple trusted applications.

Whether provided by manufacturers or academia, remote attestation remains an essential part of trusted computing solutions.
They are the foundation of trust for remote computing where the target environments are not fully trusted.
Current solutions widely differ in terms of maturity and security.
Whereas some TEEs are developed by leading processor companies and provide built-in attestation mechanisms, others still lack proper hardware attestation support and require software solutions instead.
Our study sheds some light on the limitations of \sotaAdj TEEs and identifies promising directions for future work. 

\paragraph*{Acknowledgments}
This publication incorporates results from the VEDLIoT project, which received funding from the European Union’s Horizon 2020 research and innovation programme under grant agreement No 957197, and from the Swedish Foundation for Strategic Research (SSF) aSSIsT.
 
\clearpage
\lsstyle
\printbibliography

\end{document}